\documentclass[aps,twocolumn,pra,showpacs]{revtex4}
\usepackage{amsmath,amssymb,latexsym,graphicx}

\begin{document}

\title{Time Double-Slit Interference in Tunneling Ionization}
\author{D. G. Arb\'{o}$^{\dagger }$, E. Persson, and J. Burgd\"{o}rfer}
\affiliation{Institute for Theoretical Physics, Vienna University of Technology,\\
Wiedner Hauptstra\ss e 8-10/136, A-1040 Vienna, Austria, EU}
\affiliation{$^{\dagger }$also at Institute for Astronomy and Space Physics, IAFE, CC 67,
Suc. 28 (1428) Buenos Aires, Argentina}
\date{\today}

\begin{abstract}
We show that interference phenomena plays a big role for the electron yield
in ionization of atoms by an ultra-short laser pulse. Our theoretical study
of single ionization of atoms driven by few-cycles pulses extends the
photoelectron spectrum observed in the double-slit experiment by Lindner et
al, Phys. Rev. Lett. \textbf{95}, 040401 (2005) to a complete
three-dimensional momentum picture. We show that different wave packets
corresponding to the same single electron released at different times
interfere, forming interference fringes in the two-dimensional momentum
distributions. These structures reproduced by means of \textit{ab initio}
calculations are understood within a semiclassical model.
\end{abstract}

\maketitle


\section{Introduction}

In the last years advanced laser facilities have achieved intensities of the
order of $10^{15}W/\mathrm{cm}^{2}$ and pulse lengths of the order of 10 fs,
which corresponds to few cycles of an electrical field of 800 nm wavelength 
\cite{Paulus01}. The interaction process of such short and strong pulses
with matter is a topic which has attracted much interest recently. Many
experimental (see for example \cite{Paulus94,Moshammer03}) and theoretical
studies have been performed in this line. Theoretical calculations employ
different methods: \textit{ab initio} by solving the time dependent Schr\"{o}%
dinger equation (TDSE) \cite{Dionissopoulou97,Wassaf03,Dimitriou04,arbo06})
or by using semiclassical approximation within the strong field
approximation (SFA) methods \cite{Delone91,Sand00,Milosevic03}, the
(Coulomb-) Volkov approximation \cite{Macri03,Rodriguez04}, or by
quasiclassical approximations where the electron is considered classically
but the possibility of tunneling is incorporated within the WKB approximation%
\cite{cohen01,Dimitriou04}. Very recent experiments show clear signs of
interference among different parts of the wave packet detached at different
times (different optical cycles of the electric field), when an external
linear-polarized short-laser pulse is applied \cite{Krausz05}. Clear
interference peaks in the photoelectron spectra are observed in this kind of
experiments. The photoelectron spectra of He$^{+}$ for different angles of
ejection are calculated in \cite{Chirila05}.

Another recent advance is the imaging of the momentum distribution of the
ejected electron in ionization of rare-gas atoms by a moderately strong
ultrashort laser pulse, in the transition regime from multiphoton to
tunneling ionization. In single-electron emission \cite{rudenko04}, the
longitudinal momentum distribution ($k_{z}$ along the direction of the laser
polarization) of photoelectrons from rare gases features a broad
\textquotedblleft double-hump\textquotedblright\ structure near threshold ($%
E\simeq 0$) which surprisingly resembles the $k_{z}$ distribution for
non-sequential double ionization \cite{Moshammer03}. This structure is due
to the interplay of the Coulomb interaction and laser field on the receding
trajectory of the electron \cite{Moshammer03,chen,Dimitriou04,Faisal05}. A
recently measured complex interference pattern near threshold in the
two-dimensional momentum $(k_{z},k_{\rho })$ plane \cite{rudenko04,cocke06}
was also explained as near threshold Ramsauer-Townsend diffraction
oscillations \cite{arbo06}. A simple semiclassical analysis identified the
fringes resulting from interfering paths released at different times but
reaching the same Kepler asymptote. Very recent experiments show clear signs
of interference among different parts of the wave packet detached at
different times (different optical cycles of the electric field), when an
external linear-polarized short-laser pulse is applied \cite{Krausz05}.
Non-equispaced interference peaks in the photoelectron spectra are observed
in this kind of experiments and calculations, for example the photoelectron
spectra of He$^{+}$ for different angles of ejection \cite{Chirila05}.

In this paper we generalize the study of interference in the energy (scalar)
domain \cite{Krausz05} to the momentum (vector) domain for the case of a 1-2
cycle pulse. We calculate the electron yield when a hydrogen atom is subject
to a linear polarized two-cycles laser pulse with a $\sin ^{2}$ envelope
function and show its similarity with a one-cycle pulse without envelope
function. We demonstrate that non-equispaced peaks in the photoelectron
spectrum stems from interference of different parts of the wave packets
released at different times. Firstly, we solve the TDSE, then we make some
classical considerations in order to intuitively understand the role of
classical trajectories on interference phenomena, and lastly we present a
semiclassical model which reproduce satisfactorily the calculated
electron-yield distributions.

\section{Theory and results}

We consider a hydrogen atom interacting with a linearly polarized laser
field. The Hamiltonian of the system is 
\begin{equation}
H=\frac{{\vec{p}\,}^{2}}{2}+V(r)+z\,F\,(t),  \label{hami}
\end{equation}%
where $V(r)=-1/r$ is the Coulomb potential energy, $\vec{p}$ and $\vec{r}$
are the momentum and position of the electron, respectively, and $\vec{F}(t)$
is the time dependent external field linearly polarized along the $\hat{z}$
direction. The laser pulse is chosen to be of the form 
\begin{equation}
\vec{F}(t)=\left\{ 
\begin{array}{c}
f(t)\sin (\omega t)\ \hat{z}\quad (0\leq t\leq \tau ) \\ 
\vec{0}\text{ \qquad otherwise}%
\end{array}%
\right. ,  \label{field}
\end{equation}%
where $\omega $ is the laser frequency, $\tau $ the total pulse duration,
and $f(t)$ is the envelope function. Atomic units are used throughout.

\subsection{TDSE calculations}

The time-dependent Schr\"{o}dinger equation is solved by means of the
generalized pseudo-spectral method~\cite{tong97}. The method combines a
discretization of the radial coordinate optimized for the Coulomb
singularity with quadrature methods to achieve stable long-time evolution
using a split-operator method. It allows for an accurate description of both
the unbound as well as the bound parts of the wave function $\left\vert \psi
(t)\right\rangle $. The process of detecting an electron of momentum $\vec{k}
$ can then be viewed as a projection of the wave function onto the Coulomb
wave functions \cite{Dionissopoulou97,Schoeller86,mess65}. Therefore, after
the laser pulse is turned off, the wave packet is projected onto outgoing
Coulomb wave functions which gives the asymptotic momentum distributions 
\begin{equation}
\frac{dP}{d\vec{k}}=\frac{1}{4\pi k}\left\vert \sum_{l}e^{i\delta _{l}(k)}\ 
\sqrt{2l+1}P_{l}(\cos \theta _{k})\ \left\langle k,l\right. \left\vert \psi
(\tau )\right\rangle \right\vert ^{2},  \label{coulomb}
\end{equation}%
where $\delta _{l}(k)$ is the momentum dependent Coulomb phase shift, $%
\theta _{k}$ is the angle between $\vec{k}$ and the polarization direction
of the laser field, $\widehat{z}$, $P_{l}$ is the Legendre polynomial of
degree $l$, and $\left\vert k,l\right\rangle $ is the eigenstate of the free
atomic Hamiltonian with positive eigenenergy $E=k^{2}/2$ and orbital quantum
number $l$. The Coulomb projection is needed for the cases that a physical
magnitude is not a constant of motion of the free evolution (once the
external field is turned off), i.e., the components of the kinetic momentum.
In turn, this is not the case for the photoelectron spectrum since the
energy is a constant of motion of the free evolution. Cylindrical symmetry
makes the dynamics a two dimensional problem, where the projection of the
angular momentum on the polarization direction of the laser is a constant of
motion (the magnetic quantum number $m$ is unaffected during the time
evolution). As the initial state of the system we consider the ground state
of the Hydrogen atom, ($m=0$).

Firstly, a driving field of frequency $\omega =0.05$ with envelope function $%
f(t)=F_{0}\sin ^{2}(\pi t/\tau ),$ peak field $F_{0}=0.075$ corresponding to
an intensity of $2\times 10^{14}\mathrm{W/cm}^{2}$, and time duration $\tau
=251$ (which involves two complete optical cycles) is used. The inset of
Fig. 1 (a) shows the pulse shape. The photoelectron spectrum, shown in Fig.
1 (a), has a non-equally spaced peaked structure, whose separation between
two consecutive peaks increases with energy. Recent simulations \cite%
{Chirila05} have also shown the existence of these peaks in the density
probabilities at different angles of ejection but for a much stronger field (%
$10^{16}\mathrm{Wcm}^{-2}$) applied to He$^{+}$ atoms calculations.

\includegraphics[width=7.7cm,angle=0,bb=30 130 550 800]{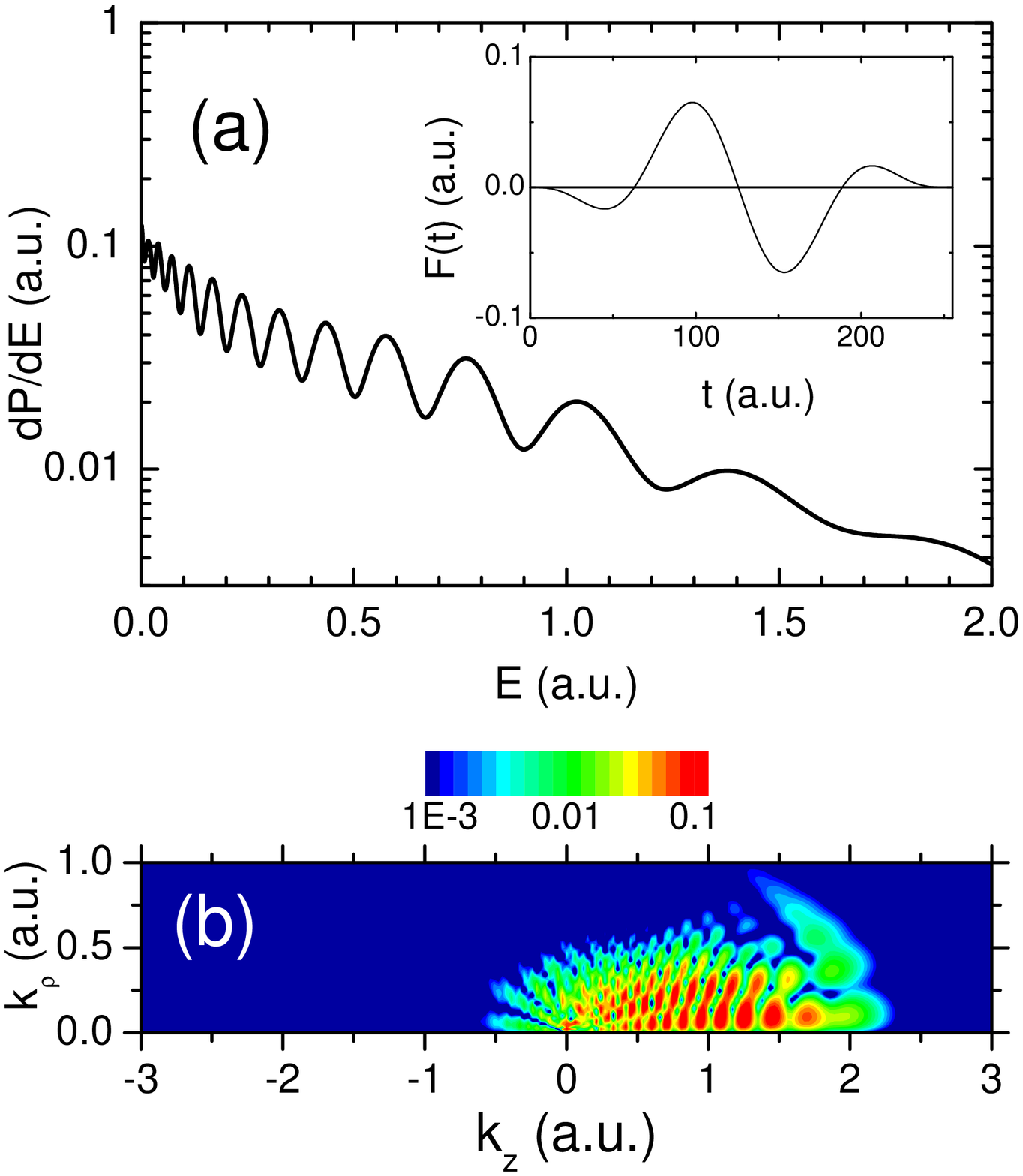}\\

{\bf FIG. 1}: (a) Photoelectron spectra for a two-cycle electric field of
frequency $\omega =0.05$ a.u., peak field $F_{0}=0.075$ a.u. and duration 
$\tau =4\pi /\omega =251.$ Inset: Electric field as a function of time. (b)
Doubly differential electron momentum distribution in cylindrical
coordinates ($k_{z},k_{\rho }$) in logaritmic scale for the same laser pulse.\\

In order to develop a more complete analysis we show in Fig. 1 (b) the
double differential momentum distribution as a function of the final
longitudinal ---$k_{z}$--- and transversal ---$k_{\rho }=\sqrt{%
k_{x}^{2}+k_{y}^{2}}$--- momentum of the electron $\frac{d^{2}P}{dk_{\rho
}dk_{z}}$. Different characteristics of the two-dimensional distribution are
worth to be highlighted: (i) about the 95\% of the distribution lies in the
region of positive longitudinal momentum ($k_{z}>0$), (ii) more specifically
the distribution is constrained to the region $-0.5\lesssim k_{z}\lesssim
2.5 $ and $k_{\rho }\lesssim 0.6,$ and the most important ,(iii) a pattern
with strip shape. It can also be observed some small radial structure for
positive $k_{z}$ near threshold. This kind of structure has recently be
understood as generalized Ramsauer-Townsend type diffraction oscillations
resulting from interfering paths released at different times \cite{arbo06}.

In the following we consider a pulse with constant envelope function $%
f(t)=F_{0}$ within $[0,\tau ]$ and $0$ outside for two different values for
the pulse duration: $\tau =\pi /\omega =63$ (half-cycle pulse) and $\tau
=2\pi /\omega =126$ (full-cycle pulse). In the case of the half-cycle pulse,
the field, obviously, does not correspond to a traveling but to a standing
wave [see Fig. 2 (a)]. We can observe the similarity of the shape of the
one-cycle pulse of Fig. 2 (a), with the two-cycle pulse with envelope
function in the inset of Fig. 1, where we can consider only one effective
cycle (at the center) while the other is rather weak. As the pulse shapes of
these two pulses are similar so should it also be the electron yield
distributions.

\includegraphics[width=7.7cm,angle=0,bb=30 150 550 730]{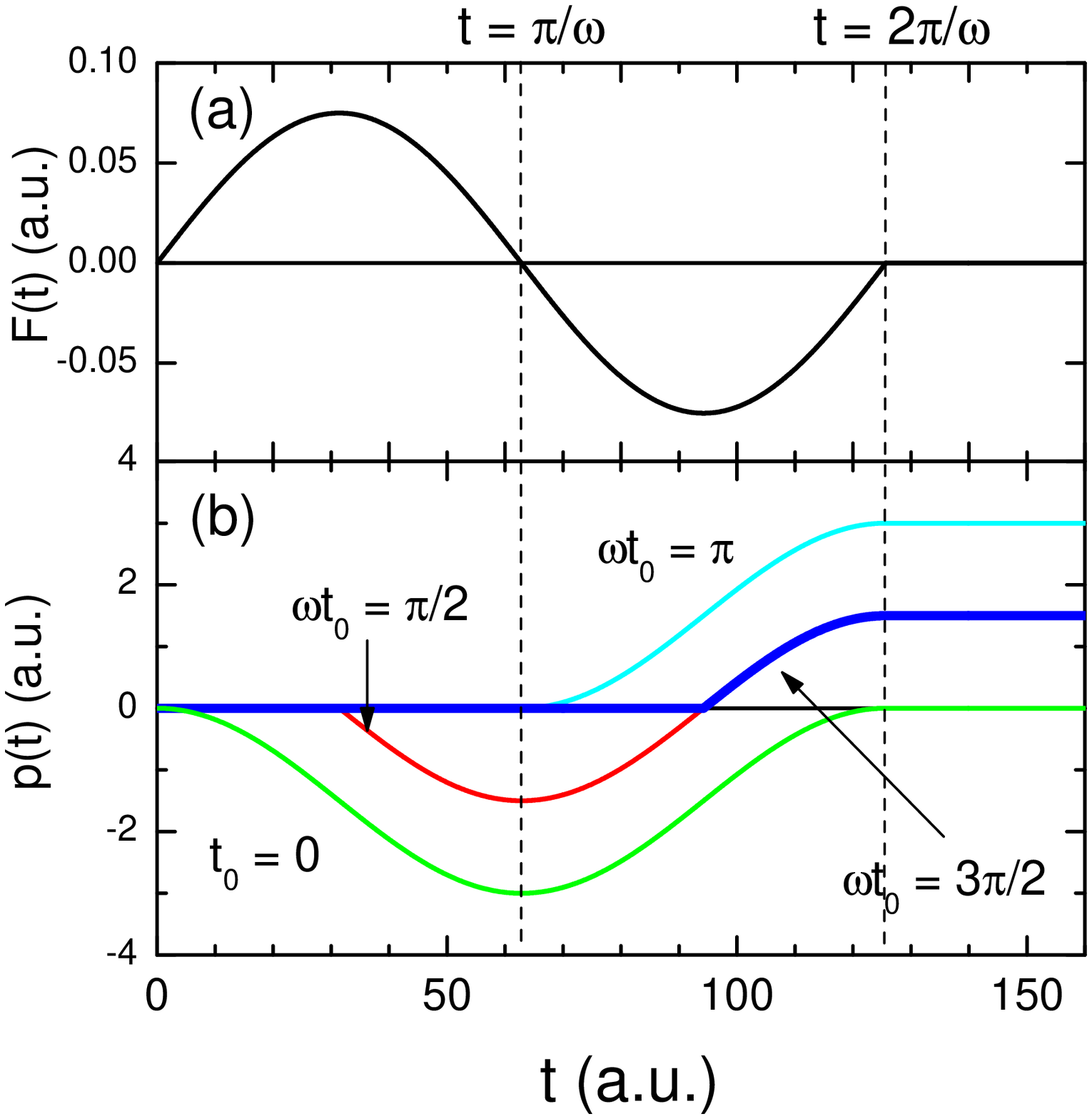}\\

{\bf FIG. 2}: (a) Electric field of frequency $\omega =0.05,$ duration $%
\tau =2\pi /\omega =126$ a.u. and amplitude $F_{0}=0.075$ a.u. as a function
of time (1 cycle and square envelope function). (b) Kinetic momentum for
different classical trajectories: electron detached at time $t_{0}=0,\pi
/2\omega ,\pi /\omega ,3\pi /2\omega ,$ and $2\pi /\omega $, as indicated.
The momentum of the trajectories $t_{0}=\pi /2\omega $ and $3\pi /2\omega $
coincide in the time interval $t\geq 3\pi /2\omega $.\\

In Fig. 3 (a) the photoelectron spectra for one-cycle pulse must be compared
to the one of Fig. 1. In Fig. 3 (a) we showed the photoelectron spectra for
a half-cycle and a full-cycle pulses are shown. The total ionization
probability, which can be calculated as the integral of the energy
distribution, is higher for a full cycle pulse than for a half cycle pulse,
as expected. After half a cycle the ionization probability is $0.06$, while
after a full cycle pulse it is the double, $0.12$. This shows that the
system is in the linear ionization regime very far away from the saturation
limit (depletion of the ground state). We will make use of this fact below,
in subsection C. Both energy-distributions feature a peak structure with
bigger separation between consecutive peaks as the energy increases. In
turn, the photoelectron spectrum of a half-cycle pulse shows no oscillating
structure. In Fig. 3 (b) the double-differential momentum distribution for a
half-cycle shows a complete smooth distribution in the negative longitudinal
momentum -$k_{z}$- region. The pulse is not only responsible for the
ionization of the atom but also for the motion of the wave packet towards
the negative longitudinal momentum region. In turn, when a second half-cycle
is included [Fig. 3 (d)] the whole wave packet moves towards the positive
region and a clear fringe structure appears. This fact leads us to think
that the peak structure of the one-cycle pulse spectrum emerges as a
consequence of an interference effect. When considering a full cycle pulse,
the wave packet created (released) during the first half cycle is pulled to
the negative longitudinal momentum region after the first half cycle, but
then is pulled towards the positive longitudinal momentum region by the
second half-cycle. We observe that while the wave packet is spread about $3$
a.u. in the positive longitudinal momentum, about only $0.5$ a.u. (17\%) in
the transversal momentum. Additionally to the ionization in the first half
cycle, another part of the wave packet appears in the continuum during the
second half cycle which is also pulled towards the positive longitudinal
momentum region. These two wave packets interfere, which can be observed in
the double differential momentum distribution as clear interference fringes
[Fig 3 (c)] and in the photoelectron spectrum as non-equispaced peaks [Fig.
3 (a)]. The spacing between the peaks increases with energy.

\includegraphics[width=7.7cm,angle=0,bb=30 150 550 780]{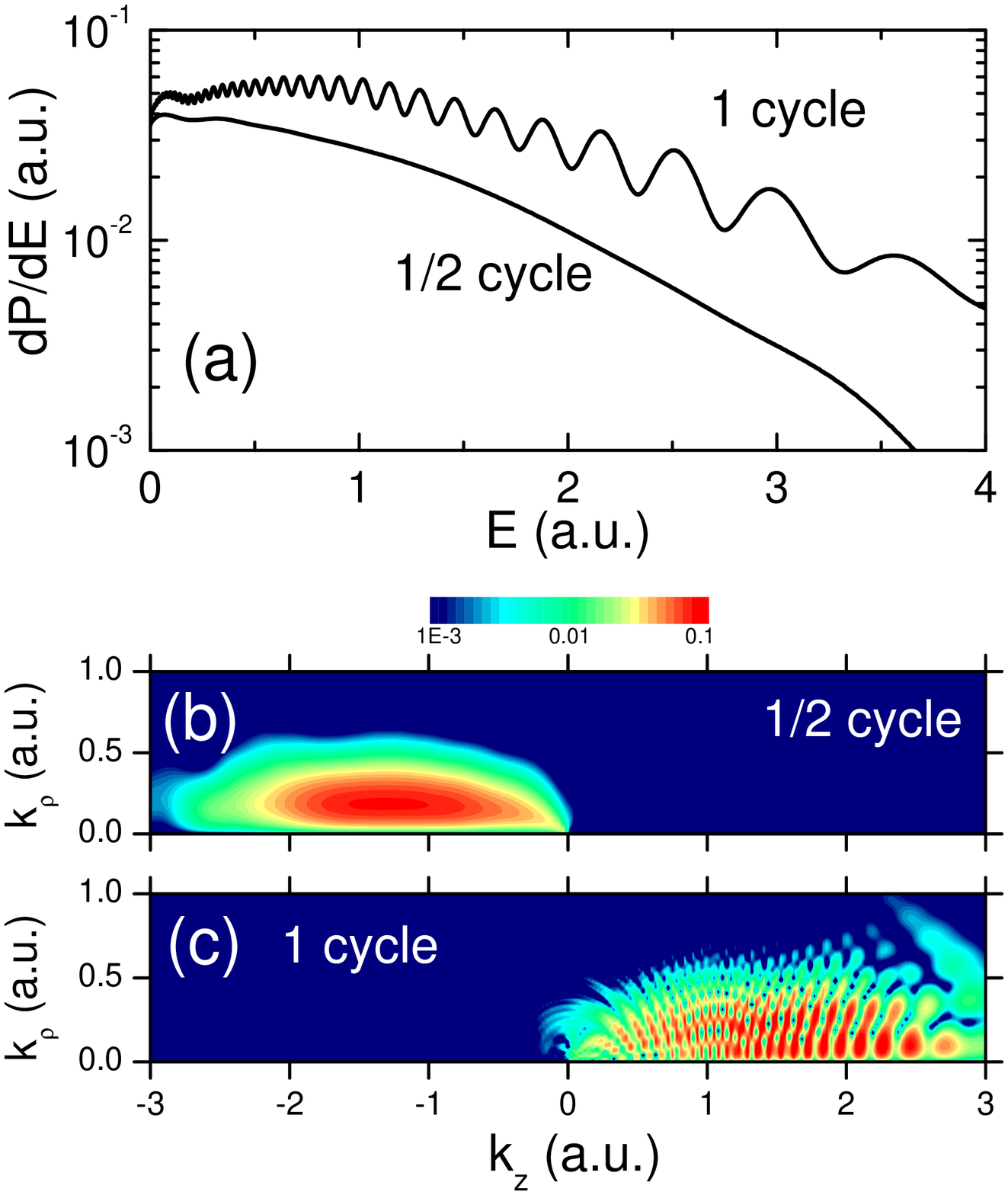}\\

{\bf FIG. 3}: (a) \textit{Ab initio} photoelectron spectra for the case
of the electric field in Fig. 2 (a): after a half-cycle pulse ($\tau =\pi
/\omega $) and one-cycle pulse ($\tau =2\pi /\omega $). (b) \textit{Ab initio%
} doubly differential electron momentum distribution after the half-cycle
pulse (one slit). (c) \textit{Ab initio} doubly differential electron
momentum distribution after the one-cycle pulse (two slits).\\

These interference processes can be reproduce with a simple semiclassical
model of two electronic wave packets escaping from the nucleus near the
extremes of the electric field and interfering in momentum space. We will
see in subsection C that the mentioned interference process can be
understood as the interference in momentum space of two electron classical
trajectories escaping from the nucleus, one at each half cycle near the
extremes of the electric field. We will give a clear analysis of this
interference effect below by means of a semiclassical approach, but before a
few classical considerations should be stated.

\subsection{Classical considerations}

In this subsection we want to intuitively understand the physics of
interference described in last subsection. We will see that it is sufficient
to use a one-dimensional picture to glimpse the interference of the ionized
atom. By using the classical equation of motion and considering the electron
to detach from the nucleus at time $t_{0}$ with zero velocity (Simple Man's
Model \cite{corkum}), it acquires the following (longitudinal) momentum 
\begin{equation}
p(t)=\frac{F_{0}}{\omega }\left( \cos \omega t-\cos \omega t_{0}\right) .
\label{clpz}
\end{equation}%
In Eq. (\ref{clpz}), we make use of the strong field approximation (SFA),
thus the effect of the atomic potential on the detached electron is
neglected. The momentum $p$ corresponds to the longitudinal momentum $k_{z}$
in the three-dimensional case.

Most probably the atom is ionized at the extremes of the one-cycle electric
field, i.e., $\omega t_{0}=\pi /2$ and $3\pi /2,$ thus Eq. (\ref{clpz})
leads to $p(t)=\left( F_{0}/\omega \right) \cos \omega t.$ Therefore, after
a half-cycle pulse ($\tau =\pi /\omega $) the momentum of the electron is $%
p(t)=-F_{0}/\omega $, and after a complete-cycle pulse ($\tau =2\pi /\omega $%
), $p(t)=F_{0}/\omega $ [see Fig. 2 (b)]. In Fig. 3 (b) and (c) we can
observe that the center of mass of the wave packet follows the classical
quiver motion with amplitude $F_{0}/\omega $, i.e., right after the first
half cycle it is negative, $\left\langle k_{z}\right\rangle \simeq
-F_{0}/\omega =-1.5$, and right after the complete cycle it is positive, $%
\left\langle k_{z}\right\rangle \simeq F_{0}/\omega =1.5$. The limit cases
are trajectories of the electron released right at the beginning (or end) of
the pulse $t_{0}=0$ (or $2\pi /\omega $) and right after the first half
cycle ($t_{0}=\pi /\omega $). From Fig. 2 (b) [and from Eq. (\ref{clpz})] it
can be easily seen that the momentum is constrained to values $%
-2F_{0}/\omega <p<0,$ right after a half-cycle pulse, and $0<p<$ $%
2F_{0}/\omega $, right after a one-cycle pulse. In Fig. 3 (b) and (c) we
observe that the quantum wave packets follow this classical constrains quite
accurately.

In Fig. 2 (b) we see an example of two different trajectories released at
times $t_{0}=\pi /2\omega $ and $3\pi /2\omega $ having the same final
momentum. It is straightforward from Eq. (\ref{clpz}) to deduce that every
pair of trajectories will have the same final momentum $p_{1}=p_{2}$, if
their respective release times $t_{1}$ and $t_{2}$ accomplish with the
relation%
\begin{equation}
\omega \left( t_{1}+t_{2}\right) =2\pi .  \label{intcond}
\end{equation}%
This is exactly the condition for interference of two classical trajectories
in the semiclassical model, as we will see in the next subsection.

In Fig. 2 we have considered only \textquotedblleft direct
electrons\textquotedblright , that is those which do not suffer any process
of rescattering. In this one-dimensional model a rescattering process means
an inversion of $\pi $ in the scattering angle (change of sign in the
momentum). Thus, the electrons would travel to the negative position
direction, $-z$, instead of being expelled towards the positive direction
after a complete cycle. As there are no other trajectories going in the $-z$
direction, so there is no possibility of interference. Therefore, next these
\textquotedblleft indirect electrons\textquotedblright\ can be neglected for
a one-cycle pulse and will be excluded from our semiclassical model of
interference. This is not the case for a longer (more than one cycle) pulse
where rescattering can occur in different cycles with the ensuing production
of interference arising from indirect electron trajectories.

\subsection{Semiclassical model}

Throughout this subsection we solve the TDSE under a set of approximations
and will arrive to a simple analytical semiclassical solution \cite%
{lewenstein}. These approximations, some of them already mentioned, are to
be introduced throughout this subsection. We consider that there is only one
atomic bound state ($\left\vert 0\right\rangle $) neglecting the bound
excited states. In order to solve the Schr\"{o}dinger equation with
Hamiltonian given by Eq. (\ref{hami}) we assume the following ansatz 
\begin{equation}
\left\vert \psi (t)\right\rangle =e^{iI_{p}t}\left[ a(t)\left\vert
0\right\rangle +\int dvb(\vec{v},t)\left\vert \vec{v}\right\rangle \right] ,
\label{ansatz}
\end{equation}%
where $\left\vert \vec{v}\right\rangle $ is an eigenstate of the continuous
spectrum with velocity $\vec{v},$ and $I_{p}$ is the atomic ionization
potential ($I_{P}=0.5$ for the hydrogen case). When we include Eq. (\ref%
{ansatz}) into the Schr\"{o}dinger equation we arrive to the following
expression for the amplitude of the continuum states: 
\begin{equation}
b(\vec{k},t)=-\sum_{i=1}^{2}\left[ \frac{2\pi iF(t_{SP}^{i})}{|\vec{k}+\vec{A%
}(t_{SP}^{i})|}\right] ^{1/2}d^{\ast }(\vec{k}+\vec{A}%
(t_{SP}^{i}))e^{iS(t_{SP}^{i},t)},  \label{bsol}
\end{equation}%
where $S$ accounts for the Volkov action, 
\begin{equation}
S(t^{\prime },t)=-\int_{t^{\prime }}^{t}dt^{\prime \prime }\left[ \frac{(%
\vec{k}+\vec{A}(t^{\prime \prime }))^{2}}{2}+I_{p}\right] ,  \label{action}
\end{equation}%
$d^{\ast }(\vec{v})=\left\langle \vec{v}\right\vert z\left\vert
0\right\rangle $ is the dipole element of the bound-continuum transition, $%
\vec{A}(t)=-\int_{0}^{t}dt^{\prime }\vec{F}(t^{\prime })$ is the vector
potential of the laser field divided by the speed of light, and $\vec{k}$ is
the canonical momentum defined as $\vec{k}=\vec{v}(t)-\vec{A}(t)$. To arrive
to Eq. (\ref{bsol}) we have made use of $\left\langle \vec{v}\right.
\left\vert \vec{v}^{\prime }\right\rangle =\delta (\vec{v}-\vec{v}^{\prime
}) $, and the following approximations: (i) we neglect the depletion of the
ground state ($a(t)=1$), (ii) continuum-continuum transitions are of the
form $\left\langle \vec{v}\right\vert z\left\vert \vec{v}^{\prime
}\right\rangle =i\nabla _{v}\delta (\vec{v}-\vec{v}^{\prime })$, and (iii)
in order to get an analytical solution we apply the saddle-point
approximation method \cite{Arfken85}. $t_{SP}^{i}$ ($i=1,2$) are the
solutions of the stationary phase action 
\begin{equation}
\frac{\partial S(t^{\prime }=t_{SP},t)}{\partial t^{\prime }}=\frac{\left[ 
\vec{k}+\vec{A}(t_{SP})\right] ^{2}}{2}+I_{p}=0.  \label{Seq0}
\end{equation}%
The solutions of Eq. (\ref{Seq0}) are complex: 
\begin{eqnarray}
t_{SP}^{1} &=&\frac{1}{\omega }\cos ^{-1}\left[ 1-\left( k_{z}\mp i\sqrt{%
2I_{p}+k_{\rho }^{2}}\right) \frac{\omega }{F_{0}}\right]  \notag \\
&&  \label{tsp} \\
t_{SP}^{2} &=&\frac{2\pi }{\omega }-t_{SP}^{1}.  \notag
\end{eqnarray}%
Therefore, the sum in Eq. (\ref{bsol}) must be understood as a two-term sum
over the two different times $t_{SP}^{1}$ and $t_{SP}^{2}$ of Eq. (\ref{tsp}%
). These two terms correspond to two different classical trajectories. The
two ionization times in Eq. (\ref{tsp}) accomplishes with the interference
condition of Eq. (\ref{intcond}).

Finally, the photoelectron spectrum can be written as $\frac{dP}{dE}=2\pi
\int_{-1}^{1}d\cos \theta _{k}\sqrt{2E}\ \left\vert b(\vec{k},t=2\pi /\omega
)\right\vert ^{2}$, since the energy is a constant of motion of the free
evolution (after the electric pulse is turned off), thus the energy
distribution is invariant when taking the asymptotic limit $t\rightarrow
\infty $. In this case, the kinetic momentum is not \textit{a priori}
invariant, but within the strong field approximation, once the electron is
detached, it is not affected any longer by the core potential and, in
consequence, the momentum distribution does not change when taking the limit 
$t\rightarrow \infty $. The momentum distribution $\frac{dP}{d\vec{k}}%
=\left\vert b(\vec{k},t=2\pi /\omega )\right\vert ^{2}$ can be written as 
\begin{equation}
\left\vert b(\vec{k},t=2\pi /\omega )\right\vert ^{2}=B(\vec{k})\cos
^{2}[\Delta S(\vec{k})]  \label{interf}
\end{equation}%
where the phase $\Delta S(\vec{k})=S(t_{SP}^{2},t)-S(t_{SP}^{1},t)$ is
responsible for the interference process, and the factor $B(\vec{k})$ is the
ionization probability at a time $t_{SP}^{i}$ and states only for the
modulation of the distribution of the unbound electrons. By making use of
the simplified version of the saddle point method, where the ionization
times of Eq. (\ref{tsp}) are considered real (Simple Man's Model), the
ionization probability can be written as \cite{Chirila05}%
\begin{equation}
B(\vec{k})=\frac{\pi ^{2}}{2(2I_{p}+k_{\rho }^{2})|F(t(\vec{k}))|^{2}}\exp %
\left[ -\frac{2(2I_{p}+k_{\rho }^{2})^{3/2}}{3|F(t(\vec{k}))|}\right] .
\label{smm}
\end{equation}%
The reader can easily check that $B(\vec{k})$ is invariant under the
interchange $t_{SP}^{1}$ $\longleftrightarrow $ $t_{SP}^{2}$ in Eq. (\ref%
{smm}), which is a direct consequence of the assumption that no depletion of
the ground state exists.

\includegraphics[width=8cm,angle=0,bb=30 200 550 950]{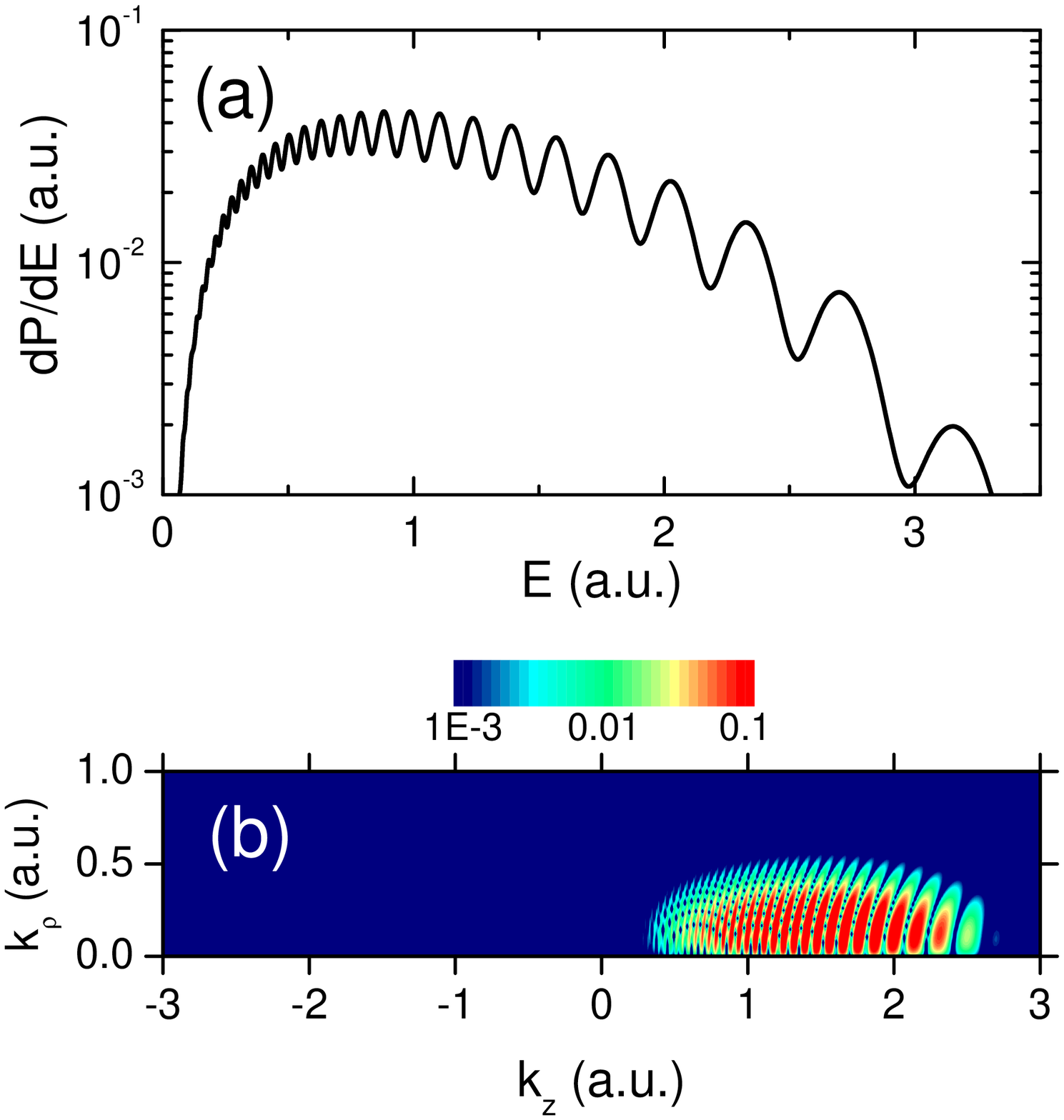}\\

{\bf FIG. 4}: (a) Semiclassical photoelectron spectrum corresponding to
the field in Fig. 2 (a). (b) Semiclassical doubly differential electron
momentum distribution after the pulse.\\

The results of the semiclassical model [Eq. (\ref{interf})] are shown in
Fig. 4. We observe a good qualitative agreement with the \textit{ab initio}
results of Fig. 3. The photoelectron spectrum calculated with the
semiclassical model in Fig. 4 (a) reproduces the oscillating structure of
the \textit{ab initio} calculations in Fig. 3 (a), but the semiclassical
model fails near threshold and in the high energy region (near classical
boundaries). The same observation can be performed for the doubly
differential momentum distributions in Fig. 4 (b) for the semiclassical
model and Fig. 3 (c) for \textit{ab initio} calculations. The interference
pattern happens to be more complicated for \textit{ab initio} calculations
even though the semiclassical model reproduces the overall interference
fringes. While the semiclassical distributions sticks (by construction) to
the classical boundaries, the \textit{ab initio} distribution invades a
little bit the classically forbidden zone.

\includegraphics[width=7.7cm,angle=0,bb=30 92 550 750]{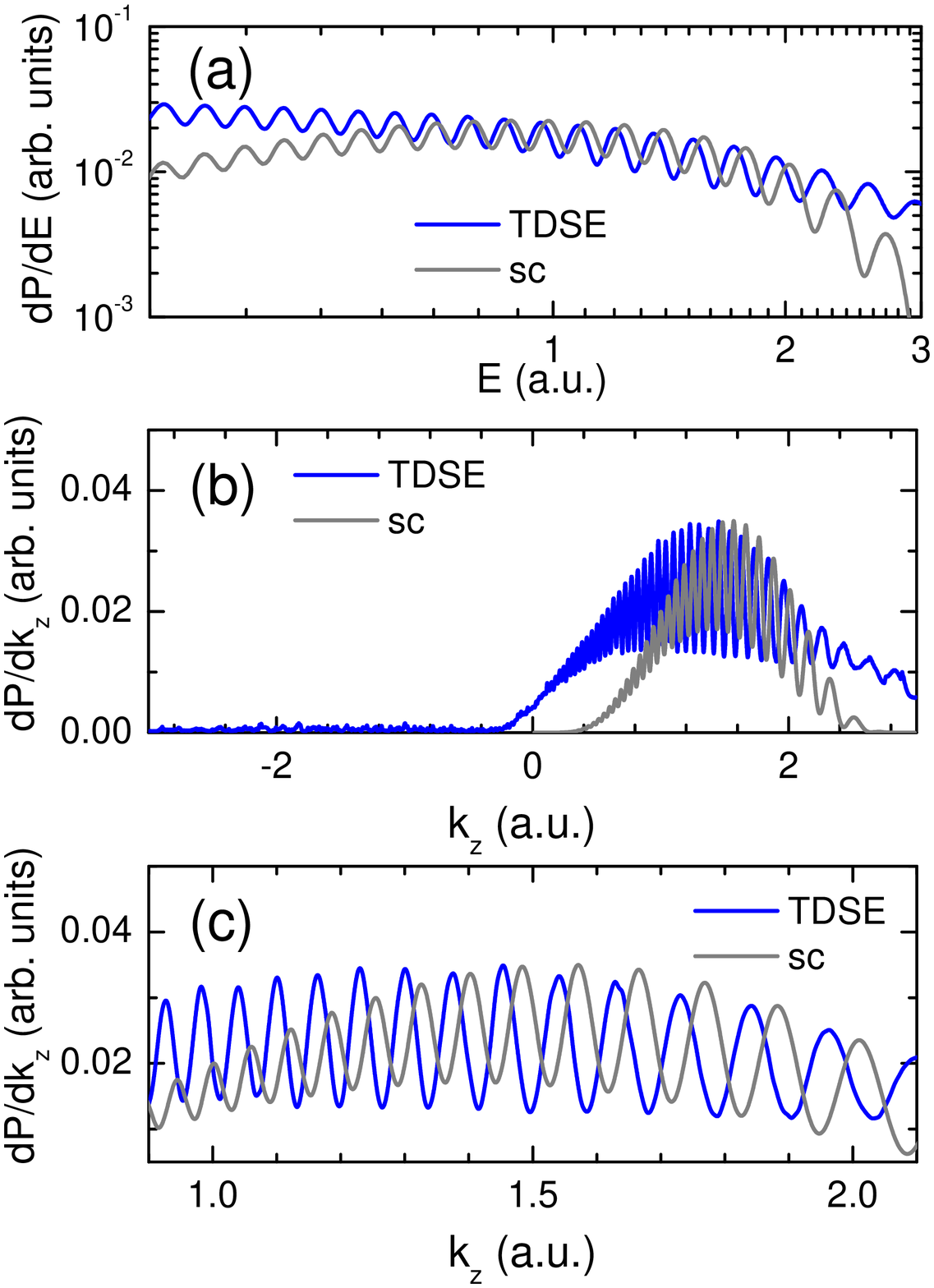}\\

{\bf FIG. 5}: (a) Photoelectron spectra by solving the TDSE and using the
semiclassical (sc) model corresponding to the field in Fig. 2 (a). (b)
Longitudinal momentum distributions by the two methods. (c) Zoom of Fig. (b) to show the agreement in separation between two consecutive peaks, even
though there is a shift.\\

In order to perform a more quantitative comparison we show in Fig. 5 the
results of both \textit{ab initio} and semiclassical calculations. For this
purpose, we have removed the excited bound states from the hydrogen spectrum
in the quantum calculation (which is present in the semiclassical model).
Therefore, the interference peaks in the photoelectron spectrum and the
momentum distribution are more enhanced since they are not masked by the
ionization coming from other bound states. Nevertheless, the distributions
of Fig. 5 are very similar to the one with the full spectrum (Fig. 3). In
Fig. 5 (a) we observed that for a long range of energies the agreement in
the position of the peaks of the semiclassical energy spectrum and the TDSE
one is very good. According to the classical mechanics [Eq. (\ref{clpz})]
the energy spectrum is constrained to $E<2\left( \frac{F_{0}}{\omega }%
\right) ^{2}=4.5$ a.u. and, as mentioned before, we observe in Fig 5 (a)
that the agreement is not so good near the extremes of the energy range,
were the semiclassical spectrum decreases more abruptly. This feature is
also observe for the longitudinal momentum distribution $\frac{dP}{dk_{z}}$
in Fig. 5 (b). In most of the range the agreement is good as it can be seen
in a zoom of the momentum distribution [Fig. 5 (c)], but it is not very good
near the borders of the classical domain $0<k_{z}<\frac{2F_{0}}{\omega }=3$
a.u., as expected according to the semiclassical theory. In both energy and
momentum distribution the envelope function of Eq. (\ref{smm}) does not
reproduce accurately the quantum results. Nevertheless, the interference
pattern of ab initio calculations can be reproduced by the semiclassical
model given by the second factor of Eq. (\ref{interf}). In figures 5 (a) and
(c) the distance between two consecutive peaks is very similar in the two
models but there is a shift in the position of the peaks of one model with
respect to the other. This shift could be due to the effect of the coulomb
potential on the ejected electron, which is present in the ab initio
calculations but is neglected in the semiclassical model (SFA).
\section{Conclusions}

In this article we have shown theoretical studies on the interference
effects observed in the electron distributions of ionized hydrogen atoms
subject to a linearly polarized short laser pulse. We have extended previous
analysis in the energy domain to a full three-dimensional one in the
momentum domain. The two-dimensional electron momentum distribution after a
full cycle pulse evidences interference fringes. We have recognize the peak
structure in the photoelectron spectrum and the longitudinal momentum
distribution as the interference phenomenon between the wave packets
released at the first and second half cycle, functioning each half cycle as
an independent slit. In the understanding of the interference phenomenon we
have made use of a simplified semiclassical model which has the advantage of
being analytical. The semiclassical model reproduces quite well the
interference patterns in the spectra of ejected electrons.

This work was supported by the SFB 016 ADLIS and the project P15025-N08 of
the FWF (Austria), and by EU project HPRI-2001-50036. D.G.A. acknowledge
support by Conicet of Argentina.


\end{document}